\documentclass[a4paper]{amsart}

\usepackage{graphics,amssymb,enumerate}
\usepackage{hyperref}
\usepackage[dvips]{graphicx}
\usepackage[all]{xy}

\def\bd{\begin{displaymath}}
\def\ed{\end{displaymath}}
\def\be{\begin{equation}}
\def\ee{\end{equation}}
\newtheorem{theorem}{Theorem}

\theoremstyle{definition}

\newtheorem{ex}{Example}

\newcommand{\g}{\mathcal{G}}

\begin{document}

\Large
\title[  Birkhoff integrable potentials]{The last integrable case of kozlov-Treshchev Birkhoff integrable  potentials}

\author{Pantelis A.~Damianou}
\address{Department of Mathematics and Statistics\\
University of Cyprus\\
P.O.~Box 20537, 1678 Nicosia\\Cyprus}
\email{damianou@ucy.ac.cy}

\author{Vassilis Papageorgiou}
\address{Department of Mathematics\\
University of Patras\\ 26500 Patras  \\Greece }
\email{vassilis@math.upatras.gr}

\thanks{Supported in part by Cyprus-Greece cooperation program  CY-GR /0603/55
}

\begin{abstract}
We establish the integrability of the last open case in the
Kozlov-Treshchev classification of Birkhoff integrable Hamiltonian
systems. The technique used  is a modification of the so called
quadratic Lax pair for $D_n$ Toda lattice combined with a method
used by M. Ranada in proving the integrability of the Sklyanin
case.
\end{abstract}

\date{November 2006}
\maketitle

\Large
\section{Introduction}

In this paper we prove the complete integrability of the
Hamiltonian system defined by

\be \label{eq1}   H = \sum_{i=1}^n \,  \frac{1}{2} \, p_i^2 + \sum
_{i=1}^{n-1} \,  e^{ q_i-q_{i+1}}  + e^{q_{n-1}+q_n} +e^{-q_1} +
e^{-2 q_1} \ . \ee

The  integrability of this system was conjectured in \cite{kozlov}
and in the book of V.V. Kozlov  \cite{kozlovbook}.  This system
appears first in the classification of Birkhoff integrable systems
by Kozlov and Treshchev \cite{kozlov}. The classification involves
systems with exponential interraction with sufficient number of
integrals, polynomial in the momenta. The classification gives
necessary conditions for a system with exponential interraction to
be Birkhoff integrable. The integrability (or not) of each system
in the list was established case by case using various techniques.
The only open case which remains is the case of system
(\ref{eq1}).  For this reason, this last case became sort of
famous and  we refer to it as the last integrable case of
Kozlov-Treshchev potentials. We now give a brief historical review
of this area, including previous progress in establishing the
integrability of   Birkhoff integrable systems.

We begin with the following  general definition which involves
systems with exponential interaction: Consider a Hamiltonian of
the form \be H=\frac{1}{2} ({\bf p}, {\bf p})+\sum_{i=1}^N e^{
({\bf v}_i\,,\, {\bf q}) }  \ , \label{a1} \ee where ${\bf
q}=(q_1, \dots, q_n)$,  ${\bf p}=(p_1, \dots, p_n)$, ${\bf v}_1,
\dots, {\bf v}_N$ are vectors in ${\bf R}^n$ and $(\ , \ )$ is the
standard inner product in ${\bf R}^n$.
 Following \cite{kozlovbook} we call the  set of vectors  $\Delta=\{ {\bf v}_1, \dots, {\bf v}_N \}$ the spectrum of the system.

Let $M$ be the $N \times N$ matrix whose elements are \bd
M_{ij}=\left( {\bf v}_i, {\bf v}_j \right) \ . \ed Hamilton's
equations  of motion can be transformed by a generalized Flaschka
transformation  to a polynomial system of $2N$ differential
equations. The transformation is defined as follows:

\be a_i=-e^{ ({\bf v}_i  , {\bf q})} ,  \qquad b_i=( {\bf v}_i,
{\bf p}) \label{a66}  \ . \ee
 We end--up with a system of polynomial differential  equations
\begin{eqnarray}
\dot{a}_k&=&a_k b_k \nonumber  \\
\dot{b}_k&=&\sum_{i=1}^N M_{ki} a_i  \label{a99} \ .
\end{eqnarray}
Equations (\ref{a99}) admit the following two integrals \be
F_1=\sum_{i=1}^N \lambda_i b_i  \ , \qquad  F_2=\prod_{i=1}^N
a_i^{\lambda_i}  \label{a98} \ee provided that  there exist
constants $\lambda_i$ such that $\sum_{i=1}^N \lambda_i {\bf
v}_i=0$. Such integrals always exist for $N>n$. One can define a
canonical bracket  on the space of variables $(a_i, b_i)$ by the
formula \bd \{b_i, a_j \}=({\bf v}_i, {\bf v}_j ) a_j \ed and all
other brackets equal to zero. The integrals $F_1$ and $F_2$ are
Casimirs of this bracket.

An interesting special case of (\ref{a1}) occurs  when  the
spectrum is a system of simple roots for a simple Lie algebra
 $\g$. In this case $N=l={\rm rank} \,  \g$. It is worth mentioning that the case where $N, n$ are arbitrary is
an open and unexplored area of research. The main exception is the
work of Kozlov and Treshchev \cite{kozlov}  where a classification
of system (\ref{a1}) is performed under the assumption that the
system possesses $n$  polynomial (in the momenta) integrals.  We
also note the papers by Ranada  \cite{ranada},   Annamalai,
Tamizhmani  \cite{anna}, Emelyanov \cite{emelyanov}, Emelyanov and
Tsygvintsev \cite{emelyanov2}.
 Such systems are called Birkhoff integrable. For each  Hamiltonian in (\ref{a1}) we
associate a Dynkin type diagram as follows: It is a graph whose
vertices correspond to the elements of  $\Delta$. Each pair of
vertices ${\bf v}_i$, ${\bf v}_j$ are connected by \bd { 4 ({\bf
v}_i, {\bf v}_j)^2 \over ({\bf v}_i,{\bf v}_i) ({\bf v}_j, {\bf
v}_j) } \ed edges.

\bigskip
\noindent \begin{ex}  The origin of systems of exponential
interraction is the classical Toda lattice which corresponds to a
Lie algebra of type $A_{n-1}$. In other words $N=l=n-1$
 and we choose $\Delta$ to
be the set: \bd {\bf v}_1=(1,-1,0, \dots, 0),   \dots \dots \dots
, {\bf v}_{n-1}=(0,0, \dots,0, 1,-1) \ . \ed The graph  is the
usual Dynkin diagram of a Lie algebra of type $A_{n-1}$. The
Hamiltonian becomes: \be
  H(q_1, \dots, q_n, \,  p_1, \dots, p_n) = \sum_{i=1}^n \,  { 1 \over 2} \, p_i^2 +
\sum _{i=1}^{n-1} \,  e^{ q_i-q_{i+1}}  \ , \label{a2} \ee which
is the well-known classical, non--periodic Toda lattice.  The
interability of this system was established in
  \cite{flaschka1},  \cite{flaschka2},  \cite{henon},  \cite{manakov},
  \cite{moser}.
\end{ex}

As we already mentioned, the  Toda lattice was generalized to the
case where the spectrum corresponds to a root space of an
arbitrary simple Lie group.
 This generalization is due to Bogoyavlensky \cite{bogo3}. These
systems were studied extensively in
 \cite{kostant} where the solution of the systems  was connected intimately with the representation
theory of simple Lie groups. There are also studies by
 Olshanetsky and Perelomov
\cite{olshanetsky} and Adler, van Moerbeke \cite{avm}. The case of
$D_n$-Toda lattice plays an important role in the present paper.

It is  more convenient to work, instead with   the space of  the
natural $(q,p)$ variables, with the Flaschka variables $(a,b)$
which are defined by:

\be
\begin{array}{lcl}
a_i& =&\frac{1}{2}e^{ \frac{1}{2} ({\bf v}_i, {\bf q})} \ \ \ \ \ i=1,2, \dots, N    \\
 b_i & = &-\frac{1}{2} p_i  \ \ \ \ \ \ \ \ \ \ \ \ \  i=1,2, \dots, n \ .     \label{a3}
\end{array}
\ee

We end--up with a new set of polynomial  equations in the
variables $(a,b)$.  One can write the equations in Lax pair form,
see for example \cite{perelomov}.
 The Lax pair ($L(t), B(t)$) in $\g$ can be described in terms
of the root system as follows:

\bd L(t)=\sum_{i=1}^l b_i(t) h_{\alpha_i} + \sum_{i=1}^l a_i(t)
(e_{\alpha_i}+e_{-\alpha_i}) \ , \ed

\bd B(t)=\sum_{i=1}^l a_i(t) (e_{\alpha_i}-e_{-\alpha_i})  \ . \ed
As usual $h_{\alpha_i}$ is an element of  a fixed Cartan
subalgebra and $e_{\alpha_i}$ is a  root vector corresponding to
the simple root $\alpha_i$. The Chevalley invariants of $\g$
provide for the constants of motion. In this paper we make use of
transformation (\ref{a3}) as opposed to transformation
(\ref{a66}).

The first important result in the search for integrable cases of
system (\ref{a1}) is due to Adler and van Moerbeke \cite{avm2}.
They considered the special case where the number of elements in
the spectrum $\Delta$ is $n+1$ (i.e.,  $N=n+1$). Furthermore, they
made the assumption that any $n$ vectors in the spectrum are
independent. Under these conditions a criterion for algebraic
integrability is that

\be {2 ({\bf v}_i,\, {\bf v}_j )  \over ({\bf v}_i,\, {\bf v}_i )
}  \label{a22} \ee should be in the set  $\{ 0, -1, -2, \dots  \}
$ for all $i\not=j$. The method of proof in \cite{avm2} is based
on the classical method of Kovalevskaya. The classification
obtained corresponds to the simple roots of graded Kac--Moody
algebras. The associated systems are the periodic Toda lattices of
Bogoyavlensky \cite{bogo3}. The complete integrability of these
systems using Lax pairs with a spectral parameter was already
established in \cite{avm}.

The next development in the study of system (\ref{a1}) is the work
of  Kozlov and Treshchev on  Birkhoff integrable systems.  A
system of the form (\ref{a1}) is called Birkhoff integrable if it
has $n$ integrals, polynomial in the momenta with coefficients of
the form \bd \sum \lambda_j e^{ <{\bf u}_j,\  {\bf q}>}\ , \ \ \
\lambda_j \in {\bf R}, \ \ \  {\bf u}_j \in {\bf R}^n   \ , \ed
whose leading homogeneous forms are almost everywhere independent.
We remark that in the definition given in the book of Kozlov
\cite{kozlovbook} there is no assumption on  involutivity of the
integrals. In \cite{kozlov} it is proved that the polynomial
integrals are in involution. For this reason, we will not deal
with the involution of the integrals; it follows from Lemma 5, p.
567 of \cite{kozlov}. The terminology has its origin in the work
of Birkhoff who studied the conditions for the existence of linear
and quadratic integrals of general Hamiltonians in  two degrees of
freedom.  A vector in $\Delta$  is called maximal if it has the
greatest possible length among all the vectors in the spectrum
having the same direction. Kozlov and Treshchev proved the
following theorem:
\begin{theorem} \label{thm:kt}
Assume that the Hamiltonian (\ref{a1}) is  Birkhoff integrable.
Let ${\bf v}_i$ be a maximal vector in $\Delta$ and assume that
the vector ${\bf v}_j \in \Delta$ is linearly independent of ${\bf
v}_i$.  Then \bd {2 ({\bf v}_i,\, {\bf v}_j )  \over ( {\bf
v}_i,\, {\bf v}_i ) } \ed lies  in the set $\{ 0, -1, -2, \dots
\}  $.

\end{theorem}

Note that the condition of the theorem is exactly the same as
condition (\ref{a22}) of Adler and van Moerbeke. Of course theorem
1 is more general since there is no restriction on the integer $N$
(the number of summands in the potential of (\ref{a1})). It turns
out, however,  that $N$ cannot be much bigger than $n$. In fact,
it follows from the classification that $N \le n+3$.
 A system of the form (\ref{a1}) is called complete if there exist no vector ${\bf v}$ such that the set
$\Delta \cup \{ {\bf v} \}$ satisfies the assumptions of
 Theorem \ref{thm:kt}. In \cite{kozlov} there is  a complete classification
of all possible Birkhoff integrable systems based on  Theorem
\ref{thm:kt}. The Dynkin type diagram of a complete, irreducible,
Birkhoff integrable Hamiltonian system is  isomorphic to one of
the following diagrams:

$\put(65,55){\circle{10}} \put(20,10){\circle{10}}
\put(50,10){\circle{10}} \put(80,10){\circle{10}}
\put(110,10){\circle{10}} \put(65,65){1} \put(20,20){1}
\put(50,20){1} \put(80,20){1} \put(110,20){1}
\put(25,10){\line(1,0){20}} \put(56.5,10){.....}
\put(85,10){\line(1,0){20}} \put(23.5,13.5){\line(1,1){37.9}}
\put(106.4,13.5){\line(-1,1){37.9}} \put(60,-20){(a)}$
$\put(190,60){\circle{10}} \put(280,60){\circle{10}}
\put(190,30){\circle{10}} \put(190,0){\circle{10}}
\put(220,30){\circle{10}} \put(250,30){\circle{10}}
\put(280,30){\circle{10}} \put(280,0){\circle{10}} \put(195,65){1}
\put(285,65){1} \put(195,35){1} \put(195,5){1} \put(225,35){1}
\put(255,35){1} \put(285,35){1} \put(285,5){1}
\put(195,30){\line(1,0){20}} \put(255,30){\line(1,0){20}}
\put(226.5,30){.....} \put(190,55){\line(0,-1){20}}
\put(190,25){\line(0,-1){20}} \put(280,55){\line(0,-1){20}}
\put(280,25){\line(0,-1){20}} \put(230,-20){(b)}$
$\put(340,0){\circle{10}} \put(370,0){\circle{10}}
\put(400,0){\circle{10}} \put(430,0){\circle{10}}
\put(460,0){\circle{10}} \put(400,30){\circle{10}}
\put(400,60){\circle{10}} \put(345,5){1} \put(375,5){1}
\put(405,5){1} \put(435,5){1} \put(465,5){1} \put(405,35){1}
\put(405,65){1} \put(345,0){\line(1,0){20}}
\put(375,0){\line(1,0){20}} \put(405,0){\line(1,0){20}}
\put(435,0){\line(1,0){20}} \put(400,55){\line(0,-1){20}}
\put(398,11){\vdots} \put(395,-20){(c)}$
$\put(20,-90){\circle{10}} \put(50,-90){\circle{10}}
\put(80,-90){\circle{10}} \put(110,-90){\circle{10}}
\put(140,-90){\circle{10}} \put(170,-90){\circle{10}}
\put(200,-90){\circle{10}} \put(110,-60){\circle{10}}
\put(25,-85){1} \put(55,-85){1} \put(85,-85){1} \put(115,-85){1}
\put(145,-85){1} \put(175,-85){1} \put(205,-85){1}
\put(115,-55){1} \put(25,-90){\line(1,0){20}}
\put(55,-90){\line(1,0){20}} \put(85,-90){\line(1,0){20}}
\put(115,-90){\line(1,0){20}} \put(145,-90){\line(1,0){20}}
\put(175,-90){\line(1,0){20}} \put(110,-65){\line(0,-1){20}}
\put(100,-120){(d)}$
$\put(250,-90){\circle{10}} \put(280,-90){\circle{10}}
\put(310,-90){\circle{10}} \put(340,-90){\circle{10}}
\put(370,-90){\circle{10}} \put(400,-90){\circle{10}}
\put(430,-90){\circle{10}} \put(460,-90){\circle{10}}
\put(400,-60){\circle{10}} \put(255,-85){1} \put(285,-85){1}
\put(315,-85){1} \put(345,-85){1} \put(375,-85){1}
\put(405,-85){1} \put(435,-85){1} \put(465,-85){1}
\put(405,-55){1} \put(255,-90){\line(1,0){20}}
\put(285,-90){\line(1,0){20}} \put(315,-90){\line(1,0){20}}
\put(345,-90){\line(1,0){20}} \put(375,-90){\line(1,0){20}}
\put(405,-90){\line(1,0){20}} \put(435,-90){\line(1,0){20}}
\put(400,-65){\line(0,-1){20}} \put(325,-120){(e)}$
$\put(20,-160){\circle{10}} \put(50,-160){\circle{10}}
\put(80,-160){\circle{10}} \put(110,-160){\circle{10}}
\put(140,-160){\circle{10}} \put(25,-155){1} \put(55,-155){1}
\put(85,-155){1} \put(115,-155){2} \put(145,-155){2}
\put(25,-160){\line(1,0){20}} \put(55,-160){\line(1,0){20}}
\put(115,-160){\line(1,0){20}} \put(84.5,-158.5){\line(1,0){21}}
\put(84.5,-161.5){\line(1,0){21}} \put(100,-190){(f)}$
$\put(270,-160){\circle{10}} \put(300,-160){\circle{10}}
\put(330,-160){\circle{10}} \put(360,-160){\circle{10}}
\put(390,-160){\circle{10}} \put(275,-155){2} \put(305,-155){2}
\put(335,-155){2} \put(365,-155){1} \put(395,-155){1}
\put(275,-160){\line(1,0){20}} \put(305,-160){\line(1,0){20}}
\put(365,-160){\line(1,0){20}} \put(334.5,-158.5){\line(1,0){21}}
\put(334.5,-161.5){\line(1,0){21}} \put(325,-190){(g)}$
$ \put(20,-240){\circle{10}} \put(20,-280){\circle{10}}
\put(70,-260){\circle{10}} \put(100,-260){\circle{10}}
\put(130,-260){\circle{10}} \put(160,-260){\circle{10}}
\put(210,-240){\circle{10}} \put(210,-280){\circle{10}}
\put(25,-235){4} \put(30,-270){1} \put(75,-255){2}
\put(105,-255){2} \put(135,-255){2} \put(170,-265){2}
\put(215,-235){4} \put(215,-275){1}
\put(75,-260.5){\line(1,0){20}} \put(135,-260){\line(1,0){20}}
\put(106.5,-260){.....} \put(18.3,-245){\line(0,-1){30.5}}
\put(21.6,-245){\line(0,-1){30.5}}
\put(208.3,-245){\line(0,-1){30.5}}
\put(211.6,-245){\line(0,-1){30.5}} \put(24,-244){\line(5,-2){41}}
\put(24,-237){\line(5,-2){45}} \put(24,-284){\line(5,2){46.5}}
\put(24,-277){\line(5,2){41}} \put(206,-244){\line(-5,-2){41}}
\put(210,-235){\line(-5,-2){48.5}}
\put(206,-284){\line(-5,2){46.5}} \put(206,-277){\line(-5,2){41}}
\put(15,-240){\line(0,-1){40}} \put(24,-244){\line(0,-1){34}}
\put(215,-240){\line(0,-1){40}} \put(206,-244){\line(0,-1){34}}
\put(100,-310){(h)}$
$ \put(450,-260){\circle{10}} \put(350,-260){\circle{10}}
\put(400,-240){\circle{10}} \put(400,-280){\circle{10}}
\put(455,-255){2} \put(345,-255){2} \put(405,-235){4}
\put(405,-295){1} \put(398.3,-245){\line(0,-1){30.5}}
\put(401.6,-245){\line(0,-1){30.5}}
\put(404,-244){\line(5,-2){41}} \put(404,-237){\line(5,-2){45}}
\put(404,-284){\line(5,2){46.5}} \put(404,-277){\line(5,2){41}}
\put(396,-244){\line(-5,-2){41}}
\put(400,-235){\line(-5,-2){48.5}}
\put(396,-284){\line(-5,2){46.5}} \put(396,-277){\line(-5,2){41}}
\put(405,-240){\line(0,-1){40}} \put(396,-244){\line(0,-1){34}}
\put(390,-310){(i)}$
$ \put(20,-380){\circle{10}} \put(20,-410){\circle{10}}
\put(20,-440){\circle{10}} \put(50,-410){\circle{10}}
\put(80,-410){\circle{10}} \put(110,-410){\circle{10}}
\put(160,-390){\circle{10}} \put(160,-430){\circle{10}}
\put(25,-375){2} \put(25,-435){2} \put(25,-405){2}
\put(55,-405){2} \put(85,-405){2} \put(120,-415){2}
\put(165,-385){4} \put(165,-425){1} \put(20,-385){\line(0,-1){20}}
\put(20,-415){\line(0,-1){20}} \put(25,-410.5){\line(1,0){20}}
\put(85,-410){\line(1,0){20}} \put(56.5,-410){.....}
\put(158.3,-395){\line(0,-1){30.5}}
\put(161.6,-395){\line(0,-1){30.5}}
\put(156,-394){\line(-5,-2){41}}
\put(160,-385){\line(-5,-2){48.5}}
\put(156,-434){\line(-5,2){46.5}} \put(156,-427){\line(-5,2){41}}
\put(165,-390){\line(0,-1){40}} \put(156,-394){\line(0,-1){34}}
\put(100,-470){(j)}$
$\put(410,-410){\circle{10}} \put(410,-380){\circle{10}}
\put(410,-440){\circle{10}} \put(330,-410){\circle{10}}
\put(360,-410){\circle{10}} \put(417,-410){4} \put(415,-375){9}
\put(415,-452){1} \put(330,-403){3} \put(360,-403){3}
\put(405,-380){\line(0,-1){60}} \put(415,-380){\line(0,-1){60}}
\put(408.3,-384.5){\line(0,-1){21}}
\put(411.6,-384.5){\line(0,-1){21}}
\put(408.3,-414.5){\line(0,-1){21}}
\put(411.6,-414.5){\line(0,-1){21}} \put(335,-410){\line(1,0){20}}
\put(365,-410){\line(1,0){40}} \put(360,-405){\line(1,0){50}}
\put(360,-415){\line(1,0){50}} \put(359.5,-404){\line(5,3){49}}
\put(365,-410){\line(5,3){43}} \put(363.25,-406){\line(5,3){41.5}}
\put(359.5,-416){\line(5,-3){49}} \put(365,-410){\line(5,-3){43}}
\put(363.25,-414){\line(5,-3){41.5}}
\qbezier(410,-385)(465,-410)(410,-435)
\qbezier(413,-381.7)(470,-410)(413,-438.4)
\qbezier(413,-378.4)(475,-410)(413,-441.7)
\qbezier(410,-375)(485,-410)(410,-445) \put(390,-470){(k)} $

\vskip 1cm

\noindent {\it Remark 1} In the list of  diagrams we have omitted
some cases that occur as sub--graphs of diagrams (a)--(k) (by
truncating one or more vertices).

\noindent {\it Remark 2} The Dynkin type diagram determines only
the angles between pairs of vectors in $\Delta$. In order to
reconstruct the ratios of lengths of vectors in $\Delta$ we assign
to the $i$th vertex a coefficient proportional to the square of
the length  of ${\bf v}_i$.  This explains the numbers appearing
on  the vertices of the diagrams.

We have to stress that this classification gives only necessary
conditions for a system of type (\ref{a1}) to be Birkhoff
integrable. The integrability for each system in the list should
be established case by case.  As we already mentioned, the
integrability of systems (a)--(g) was established in \cite{avm},
\cite{bogo3}. The solution of these generalized periodic Toda
lattices (associated with affine Lie algebras) was obtained by
Goodman and Wallach in \cite{goodman}. The graph (i) corresponds
to a Hamiltonian system in two degrees of freedom with potential
\bd e^{q_1}+e^{q_2}+e^{-q_1-q_2}+e^{ -\left( {q_1+ q_2 \over 2}
\right)} \ . \ed
 The additional integral can be found in
\cite{kozlov}.

Sklyanin \cite{sklyanin} pointed out  another integrable
generalization of the Toda lattice: \be
  H(q_1, \dots, q_n, \,  p_1, \dots, p_n) = \sum_{i=1}^n \,  { 1 \over 2} \, p_i^2 +
\sum _{i=1}^{n-1} \,  e^{ q_i-q_{i+1}}  + \alpha_1 e^{q_1}+
\beta_1 e^{2 q_1} + \alpha_n e^{-q_n}+\beta_n e^{-2 q_n}
 \ .  \label{a4}
\ee  He obtained this system  by means of the quantum inverse
scattering R-matrix method.  This corresponds to  system $(h)$
with associated  Hamiltonian (\ref{a4}). The case $n=2$
corresponds to the potential \bd V=e^{q_1-q_2}+ c_1 e^{2 q_2}+c_2
e^{q_2}+ c_3^{-q_1} + c_4 e^{-2 q_1} \ . \ed Annamalai and
Tamizhmani \cite{anna} demonstrated the integrability of this
particular case by using Noether's theorem. The second integral is
of fourth degree in the momenta. A special case of (\ref{a4}) was
considered in \cite{damianou4} using the generalization of KM
(Volterra) system due to Bogoyavlensky. The case $n=3$ (as well as
the general case) is  treated in Ranada \cite{ranada}. Ranada
proved integrability by using  a Lax pair approach. The additional
integrals are of degree 4 and 6. Ranada's approach will be used in
proving the integrability of system (\ref{eq1}). It is generally
believed that system (k) in non-integrable. In fact the arguments
in \cite{kozlov} support the non-integrability of system (k). This
leaves only system (j) which is treated in the present paper.

The only essential progress to our knowledge in proving the
integrability of system (\ref{eq1}) are the two papers
\cite{emelyanov} and \cite{emelyanov2}.
 The case $n=4$ was
studied in \cite{emelyanov2}. Based on calculations of the
Kovalevskaya exponents it was conjectured that the degrees of
homogeneity of the additional first integrals are $4,6,8$. Our
results confirm that prediction. We reproduce here a small  part
of the  table from \cite{emelyanov2}  which  contains the
Kovalevskaya exponents corresponding to the various values of the
indicial locus $c$:

\[
\begin{tabular}{|c||l|}
  \hline
  Vector ${\bf c} $ & KE exponents \\
  \hline
  \cline{1-2}
  (0,6,0,10,6,0)            & -5,-3,-2,-1, 1,1,1,2,4,6,8,8  \\
    (3,4,3,3,0,0)              & -3,-2,-1, 1, 1,1,2,3,4,5,6,8  \\
      (7,12,0,15,0,16)            & -7,-5,-3,-1, 1,1,1,2,4,6,8,14 \\
       (8,14,0,18,10,0)           & -7,-5,-3,-2,-1,1,1,2,4,6,8,16 \\
  (1,0,1,4,3,0)              & -3,-2,-1,-1,-1,1,1,2,2,2,4,8  \\
  (3,4,3,0,1,0)              & -3,-2,-2,-1,-1,1,1,2,2,3,4,8  \\
  (7,14,0,18,10,0)           & -7,-5,-3,-2,-1,1,1,2,4,6,8,16 \\
    \hline
\end{tabular}
\]

In a brief communication by Emelyanov  \cite{emelyanov} it was
reported (without proof) that the system (\ref{eq1}) was
integrable for $n=4$. In fact a few leading terms of the integrals
were displayed but our leading terms of the integrals do not match
exactly  those results. However, it is possible that the integrals
calculated by the author of \cite{emelyanov} are combinations of
the integrals of the present paper.  To quote from
\cite{emelyanov}:   ``A study of Kovalewski indices shows that it
is possible to assume for $n=5$ the degrees of additional
integrals are $4,6,8,10$ etc. If integrability is established for
an arbitrary $n$, then one can state that  only systems whose
Dynkin schemes are isomorphic to those of Fig. 1 except for scheme
(k) are Birkhoff integrable''.

\section{$D_n$ Toda systems}

\smallskip

Since the Lax pair of system (\ref{eq1})  uses in an essential way
the Lax pair of the $D_n$ Toda lattice we include a review of this
Bogoyavlensky-Toda system following \cite{damianou1},
\cite{damianou2}, \cite{damianou3}.

 The Hamiltonian for  the $D_n$ Toda lattice is

\begin{equation}
 H=   { 1 \over 2} \sum_1^n p_j^2 + e^{ q_1- q_2} + \cdots + e^{ q_{n-1}-q_n}
   + e^{q_{n-1} +q_n} \ \ \ \ \ \ \ \ \ \ \ n\ge 4 \ .
  \label{29}
\end{equation}

 We make a  Flaschka-type  transformation, $F: {\bf R}^{2n} \to {\bf
R}^{2n}$ defined by

\begin{displaymath}
 F:  (q_1, \dots, q_n, p_1, \dots, p_n) \to (a_1,  \dots, a_{n}, b_1,
\dots, b_n) \ ,
\end{displaymath}
with

\begin{equation}
  a_i  = {1 \over 2} \,  e^{ {1 \over 2} (q_i - q_{i+1} ) } \ ,   \ \ \ \ \
\ \ \ \ \  i=1,2, \dots, n-1, \ \ \ \ \ \  a_n= { 1 \over 2} e^{ {
1 \over 2}(q_{n-1}+ q_n) } \ , \label{30}
\end{equation}

\begin{displaymath}
             b_i  = -{ 1 \over 2} p_i,   \ \ \ \ \ \ \ \ \ \ i=1,2,\dots, n \ .
\end{displaymath}

\smallskip
\noindent Then

\begin{equation}
\begin{array}{lcl}
 \dot a _i& = & a_i \,  (b_{i+1} -b_i )  \ \ \ \ \ \   \ \ \ i=1,2,  \dots, n-1 \\
  \dot a_n& =& -a_n(b_{n-1}+ b_n)   \\
   \dot b _i &= & 2 \, ( a_i^2 - a_{i-1}^2 ) \ \ \ \ \ \ \ \ \ \ i=1,2,  \dots, n-2 \  {\rm and}\  i=n  \\
    \dot b_{n-1} &=& 2 (a_n^2+a_{n-1}^2-a_{n-2}^2)  \ .
\end{array} \label{d4eq}
\end{equation}

\smallskip
\noindent These equations can be written as a Lax pair  $\dot L =
[B, L] $, where $L$ is the  symmetric  matrix

\begin{equation}
  \begin{pmatrix}  b_1 &  a_1 &  & &  &    &    & \cr
                   a_1 &  \ddots  & \ddots  &   & &&   &  \cr
                    & \ddots & \ddots& a_{n-1} & -a_n &0 &   &  \cr
                   &  & a_{n-1} & b_n & 0 & a_n & &  \cr
                     &  &- a_n & 0 &  -b_n &-a_{n-1} & &\cr
                    & &0 &a_n & -a_{n-1} & \ddots & \ddots &  \cr
                    &  &&&& \ddots & \ddots & -a_1  \cr
                       &&&&& & -a_1 & -b_1       \end{pmatrix}  \ , \label{32}
\end{equation}
and $B$ is the skew-symmetric part of $L$ (In the decomposition,
lower Borel plus skew-symmetric).
\smallskip

The mapping  $F: {\bf R}^{2n} \to {\bf R}^{2n}$, $(q_i ,p_i ) \to
(a_i ,b_i )$, defined by (\ref{30}), transforms the standard
symplectic bracket   into  another symplectic bracket $\pi_1 $
given (up to a constant multiple) by

\bigskip
\noindent
\begin{equation}
\begin{array}{lcl}
\{ a_i, b_i \} & = &-{ 1 \over 2}a_i \ \ \ \ \ \ i=1,2,  \dots ,n \\
 \{ a_i,b_{i+1} \}& =& {1 \over 2} a_i \ \ \ \ \ \ \ \ i=1,2,  \dots , n-1 \\
  \{a_n, b_{n-1} \}&=& -{ 1 \over 2} a_n.    \label{d4br}
\end{array}
\end{equation}

\smallskip

We obtain a hierarchy of  invariant polynomials,  which we denote
by
\begin{displaymath}
  H_2, \  H_4, \  \dots, \ H_{2n}, \dots
\end{displaymath}
 defined by $H_{2i} = { 1 \over 2i} \  { \rm Tr} \ L^{2i}  $.
 The degrees of the first $n-1$ (independent) polynomials are $2,4, \dots,
2n-2$. We also define
\begin{displaymath}
P_n=\sqrt{ {\rm det}\,L}  \ .
\end{displaymath}
The degree of $P_n$ is $n$.  The set $ \{ H_2, \  H_4, \  \dots, \
H_{2n-2}, P_n \}$ corresponds to the Chevalley invariants for a
Lie group of type $D_n$.

Taking \bd H_2={1\over 2}  {\rm Tr} \ L^2=\sum_{i=1}^n b_i^2+ 2
\sum_{i=1}^n a_i^2 \ed
 as the Hamiltonian we have that \bd \pi_1 \nabla
H_2 \ed gives precisely equations (\ref{d4eq}).

\section{quadratic Lax pairs}

We employ the method used by Ranada in \cite{ranada}. The key idea
is the following: If $\left( L, B \right)$ is a Lax pair, so is
$\left( L^2, B \right)$. This follows easily from

\bd {d \over dt} L^2 =[B,L]L+L[B,L] =(BL-LB)L+L(BL-LB)=BL^2-L^2
B=[B, L^2]  \ . \ed

In the case of $D_4$ Toda lattice,  $L^2$ is the matrix

\small
\begin{equation}
  \begin{pmatrix}  a_1^2+b_1^2 &  a_1 ( b_1+ b_2)& a_1 a_2 & 0& 0 & 0   & 0   & 0\cr
                   a_1 (b_1+  b_2 )& a_1^2+a_2^2+b_2^2 & a_2 (b_2+ b_3)  & a_2 a_3  &-a_2 a_4  &0&  0 & 0 \cr
                    a_1 a_2& a_2 (b_2+b_3) & a_2^2+a_3^2+a_4^2+b_3^2& a_3 (b_3+ b_4) & a_4(b_4-b_3) &2 a_3 a_4 &0   & 0 \cr
                   0 & a_2 a_3 &  a_3(b_3+b_4) &  a_3^2+a_4^2+b_4^2 & -2 a_3 a_4  &a_4(b_4-b_3) & -a_2 a_4 & 0 \cr
                   0& -a_2 a_4&a_4(b_4-b_3) & -2 a_3 a_4  & a_3^2+a_4^2+b_4^2 & a_3 (b_3+b_4) & a_2
                   a_3 &0  \cr
                    0 & 0 & 2 a_3 a_4 & a_4 (b_4-b_3) &  a_3(b_3+b_4) & a_2^2+a_3^2+a_4^2+b_3^2 & a_2(b_2+b_3)&a_1 a_2\cr
                   0 & 0&0 & -a_2 a_4 & a_2 a_3 & a_2 (b_2+b_3) & a_1^2+a_2^2+b_2^2 & a_1 (b_1+b_2) \cr
                   0 & 0 &0&0& 0 & a_1 a_2 &  a_1(b_1+b_2)&                   a_1^2+b_1^2
                            \end{pmatrix}  \  .
\end{equation}

and
\begin{equation}
 B= \begin{pmatrix}  0 &  a_1 & 0 & 0& 0 & 0   & 0   & 0\cr
                    -a_1& 0 & a_2   & 0  &0  &0&  0 & 0 \cr
                    0& -a_2 & 0& a_3  & -a_4 &0 &0   & 0 \cr
                   0 & 0 &  -a_3 &  0 & 0  &a_4 & 0 & 0 \cr
                   0& 0&a_4 & 0 & 0 & -a_3 & 0 & 0  \cr
                    0 & 0 & 0 & -a_4 &  a_3 & 0& -a_2&0\cr
                   0 & 0&0 & 0 & 0& a_2  & 0 & -a_1  \cr
                   0 & 0 &0&0& 0 & 0 &  a_1&   0
                            \end{pmatrix}  \  .
\end{equation}

\Large
 Note that the Lax equations $\dot L^2=[B, L^2]$  gives the
same equations (\ref{d4eq}) plus some consistency conditions.

\section{The  case  $n=4$.}

We now consider the special case of (\ref{eq1}) for $n=4$. The
general case follows easily from this special case.   Let  \be  H
= \sum_{i=1}^4 \, { 1 \over 2} \, p_i^2 + \sum _{i=1}^3 \, e^{
q_i-q_{i+1}} + e^{q_{3}+q_4} +e^{-q_1} + e^{-2 q_1} \ . \ee

 We make a  Flaschka-type  transformation, $F: {\bf R}^{8} \to {\bf
R}^{9}$ defined by

\begin{displaymath}
 F:  (q_1, \dots, q_4, p_1, \dots, p_4) \to (a_1,  \dots, a_{5}, b_1,
\dots, b_4) \ ,
\end{displaymath}
with

\begin{equation}
  a_i  = \frac{1}{2}  \,  e^{ \frac{1} {2} (q_i - q_{i+1} ) } \ ,   \ \ \ \ \
\ \ \ \ \  i=1,2, 3, \ \ \ \ \ \  a_4= \frac{1} {2}  e^{ \frac{1}
{2} (q_{3}+ q_4) } \ , \ \ \ \ \ \ a_5=\frac{1}{\sqrt{ 2}} e^{
-\frac{1}{2}  q_1 } \ ,  \label{tran}
\end{equation}

\begin{displaymath}
             b_i  = -{ 1 \over 2} p_i,   \ \ \ \ \ \ \ \ \ \ i=1,2,3, 4 \ .
\end{displaymath}

We obtain the following equations of motion:

\begin{equation}
\begin{array}{lcl}
 \dot a _i& = & a_i \,  (b_{i+1} -b_i )  \ \ \ \ \ \   \ \ \ i=1,2, 3 \\
  \dot a_4& =& -a_4(b_{3}+ b_4)   \\
  \dot a_5 &= & a_5 b_1 \\
  \dot b_1 &=& 2 a_1^2-a_5^2-4 a_5^4 \\
   \dot b _i &= & 2 \, ( a_i^2 - a_{i-1}^2 ) \ \ \ \ \ \ \ \ \ \ i=2 \  {\rm and}\  i=4  \\
    \dot b_{3} &=& 2 (a_4^2+a_{3}^2-a_{2}^2)  \ .
\end{array} \label{kteq}
\end{equation}

 Note that the Hamiltonian in the new variables takes the form
 \be H= \sum_{i=1}^4 b_i^2+ 2 \sum_{i=1}^4 a_i^2 +  a_5^2+2
 a_5^4 \ . \label{ktham} \ee

 The image of the symplectic bracket  is now the following extension of bracket  (\ref{d4br})
 in the new
phase space $(a_1, \dots, a_5, b_1, \dots, b_4)$.

\noindent
\begin{equation}
\begin{array}{lcl}
\{ a_i, b_i \} & = &-{ 1 \over 2}a_i \ \ \ \ \ \ i=1,2,  \dots ,4 \\
\{a_5, b_1\} &=& {1 \over 2} a_5 \\
 \{ a_i,b_{i+1} \}& =& {1 \over 2} a_i \ \ \ \ \ \ \ \ i=1,2,  \dots , 3\\
  \{a_4, b_{3} \}&=& -{ 1 \over 2} a_4.    \label{ktbr}
\end{array}
\end{equation}

We denote this bracket by $w_1$.  Using the Hamiltonian
(\ref{ktham}) and the above bracket  $w_1$  gives equations
(\ref{kteq}) as is easily checked.

We now present a Lax pair for  equations (\ref{kteq}).  The method
of obtaining this Lax pair is based in the ideas of \cite{ranada}
and we omit the details.

The Lax pair has the form $\dot A=[C, A]$ where the matrices $C$
and $A$ are perturbations of the matrices $B$ and $L^2$
respectively. More precisely, $A$ is the same as $L^2$ except that
\bd A_{11}=a_1^2+b_1^2+a_5^2+2 a_5^4=A_{88}\ ,  \ed  \bd
A_{12}=a_1(b_1+b_2+\sqrt{2} i a_5^2)= A_{87}, \ \  \ \ \  i^2=-1
\ed \bd A_{21}=a_1(b_1+b_2-\sqrt{2} i a_5^2)= A_{78} \ . \ed

On the other hand $C$ differs from $B$ only at two diagonal
positions. It is the same as $B$ except that \bd C_{11}=\sqrt{2} i
a_5^2 =C_{88} \ . \ed

It is a simple calculation to show that  the equations  $\dot
A=[C, A]$ are a matrix form of equations (\ref{kteq}).

 Define  $h_{2i} = { 1 \over 2} \  { \rm Tr} \ A^{i} \ \ i=1,2,3,4 $.

\bd h_2={1\over 2} {\rm Tr} \ A =H=\sum_{i=1}^4 b_i^2+ 2
\sum_{i=1}^4 a_i^2 + a_5^2+2
 a_5^4 \ .  \ed

Let \bd \begin{array}{lcl} t_1 &=& a_1^2+{1 \over 2} a_5^2 +a_5^4 \\
t_2 &=& a_1^2 +a_2^2 \\
t_3 &=& a_2^2+a_3^2+a_4^2 \\
t_4 &=& a_3^2+a_4^2 \  . \end{array} \ed

Then  $h_4={1 \over 2} {\rm Tr} \ A^2 $
is given by \begin{align*}  h_4=b_1^4+b_2^4+b_3^4+b_4^4+ \sum_{i=1}^4 4 t_i b_i^2 +\\
4 a_1^2 b_1 b_2+ 4 a_2^2 b_2 b_3 + 4 (a_3^2-a_4^2) b_3 b_4+ s(a_1,
a_2, a_3,a_4, a_5) \ ,
\end{align*}

where \bd s= 4 a_1^2 a_2^2+4 a_2^2 a_4^2+4 a_2^2 a_3^2+12 a_3^2
a_4^2 + 8 a_1^2 a_5^4+ 2 a_1^2 a_5^2+ 2 a_3^4+ 2 a_4^4 +2 a_1^4 +2
a_2^4+ a_5^4+ 4 a_5^6+4 a_5^8 \ . \ed

Similarly $h_6$ has the form $ \sum_{i=1}^4 b_i^6+ \sum_{i=1}^4 12
t_i b_i^4 + $ other terms even in the momenta.

Finally $h_8$ has the form $ \sum_{i=1}^4 b_i^8+ \sum_{i=1}^4 8
t_i b_i^6 +$ other terms even in the momenta.

The set of functions $\{h_2, h_4, h_6, h_8 \}$ provide a maximal
set of independent integrals in involution. The fact that the
leading terms of the integrals are functions of the momenta only,
agrees with the general facts proved in \cite{kozlov}.

\section{The general case}

The procedure for the general case is very similar to the case
$n=4$.  There is no interaction between $q_1$ and the newly
introduced variables, therefore the procedure is quite identical.

Consider the Hamiltonian

 \bd    H = \sum_{i=1}^n \,
\frac{1}{2} \, p_i^2 + \sum _{i=1}^{n-1} \,  e^{ q_i-q_{i+1}}  +
e^{q_{n-1}+q_n} +e^{-q_1} + e^{-2 q_1} \ . \ed

 We make a  Flaschka-type  transformation, $F: {\bf R}^{n} \to {\bf
R}^{n+1}$ defined by

\begin{displaymath}
 F:  (q_1, \dots, q_n, p_1, \dots, p_n) \to (a_1,  \dots, a_{n+1}, b_1,
\dots, b_n) \ ,
\end{displaymath}
with

\be \label{fln}
  a_i  = {1 \over 2} \,  e^{ {1 \over 2} (q_i - q_{i+1} ) } \ ,   \ \ \ \ \
\ \ \ \ \  i=1,2, \dots, n-1, \ \ \ \ \ \  a_n= { 1 \over 2} e^{ {
1 \over 2}(q_{n-1}+ q_n) } \ ,\ \ \ \ \ \ a_{n+1}=\frac{1}{\sqrt{
2}} e^{ -\frac{1}{2}  q_1 } \ , \ee

\begin{displaymath}
             b_i  = -{ 1 \over 2} p_i,   \ \ \ \ \ \ \ \ \ \ i=1,2, \dots n \ .
\end{displaymath}

We obtain the following equations of motion:

\be
\begin{array}{lcl}
 \dot a _i& = & a_i \,  (b_{i+1} -b_i )  \ \ \ \ \ \   \ \ \ i=1,2,  \dots, n-1 \\
  \dot a_n& =& -a_n(b_{n-1}+ b_n)   \\
  \dot a_{n+1}&=& a_{n+1} b_1 \\
  \dot b_1 &=& 2 a_1^2 -a_{n+1}^2 -4 a_{n+1}^2 \\
   \dot b _i &= & 2 \, ( a_i^2 - a_{i-1}^2 ) \ \ \ \ \ \ \ \ \ \ i=2,3,   \dots, n-2 \  {\rm and}\  i=n  \\
    \dot b_{n-1} &=& 2 (a_n^2+a_{n-1}^2-a_{n-2}^2)  \ .
    \label{eqgen}
\end{array}  \ee

 Note that the Hamiltonian in the new variables takes the form
 \be H= \sum_{i=1}^n b_i^2+ 2 \sum_{i=1}^n a_i^2 +  a_{n+1}^2+2
 a_{n+1}^4 \ . \label{genham} \ee

 The image of the symplectic bracket  is now the following extension of bracket  (\ref{d4br})
 in the new
phase space $(a_1, \dots, a_{n+1}, b_1, \dots, b_n)$.

\noindent \bd
\begin{array}{lcl}
\{ a_i, b_i \} & = &-{ 1 \over 2}a_i \ \ \ \ \ \ i=1,2,  \dots ,n \\
\{a_{n+1}, b_1\} &=& {1 \over 2} a_{n+1} \\
 \{ a_i,b_{i+1} \}& =& {1 \over 2} a_i \ \ \ \ \ \ \ \ i=1,2,  \dots , n-1\\
  \{a_n, b_{n-1} \}&=& -{ 1 \over 2} a_n.
\end{array}
\ed

We denote this bracket by $w_1$.  Using the Hamiltonian
(\ref{genham}) and the above bracket  $w_1$  gives equations
(\ref{eqgen}) as is easily checked.

We now obtain a Lax pair for  equations (\ref{eqgen}).

The Lax pair has the form $\dot A=[C, A]$ where the matrices $C$
and $A$ are perturbations of the matrices $B$ and $L^2$
respectively. More precisely, $A$ is the same as $L^2$ except that
\bd A_{11}=a_1^2+b_1^2+a_{n+1}^2+2 a_{n+1}^4=A_{nn}\ ,  \ed  \bd
A_{12}=a_1(b_1+b_2+\sqrt{2} i a_{n+1}^2)= A_{(n+1)n}, \ \  \ \ \
i^2=-1, \ed \bd A_{21}=a_1(b_1+b_2-\sqrt{2} i a_{n+1}^2)=
A_{n(n+1)} \ . \ed

On the other hand $C$ differs from $B$ only at two diagonal
positions. It is the same as $B$ except that \bd C_{11}=\sqrt{2} i
a_{n+1}^2 =C_{nn} \ . \ed

It is a simple calculation to show that  the equations  $\dot
A=[C, A]$ are a matrix form of equations (\ref{eqgen}).

 Define  $h_{2i} = { 1 \over 2} \  { \rm Tr} \ A^{i} \ \ i=1,2, \dots,n $.

e.g.,  \bd h_2={1\over 2} {\rm Tr} \ A =H=\sum_{i=1}^n b_i^2+ 2
\sum_{i=1}^n a_i^2 + a_{n+1}^2+2
 a_{n+1}^4 \ ,  \ed
and $h_{2i}$ is a homogenous polynomial in the $b_i$ of degree
$2i$.

\end{document}